\documentclass[onecolumn,12pt]{article}
\usepackage{graphicx}
\begin{document}
\title{A genetic algorithm to design Laue lenses with optimal performance for focusing hard X- and $\gamma$-rays}

\author{Riccardo Camattari, Vincenzo Guidi \\
Department of Physics and Earth Sciences and INFN-Ferrara,\\ University of Ferrara, Via Saragat 1/c, 44122 Ferrara}

\maketitle

\begin{abstract}
In order to focus hard X- and $\gamma$-rays it is possible to make use of a Laue lens as a concentrator. With this optical tool it would be possible to improve the detection of radiation for several applications, spanning from the observation of the most violent phenomena in the sky to nuclear medicine applications, for diagnostic and therapeutic purposes. A code named \emph{LaueGen}, based on a genetic algorithm and aimed to designing optimized Laue lenses, has been implemented. The genetic algorithm was selected because the optimization of a Laue lens is a complex and discretized problem. The output of the code consists in the design of a Laue lens composed of diffracting crystals selected and arranged in such a way to maximize the performance of the lens. The code allows one to manage crystals of any material and crystallographic orientation. The program is structured in such a way that the user can control all the initial parameters of the lens. As a result, \emph{LaueGen} is highly versatile and can be used for the design of very small lens, e.g. for nuclear medicine, to very large lens, e.g. for satellite-borne astrophysical missions.\\

\end{abstract}

\section{Introduction}

The focalization of hard X- and $\gamma$-rays in the 100-1000 keV energy range is a topic of growing interest, because of the wealth of physical experiments that could be performed and the technological spin-off that could be derived. In fact, the observation of the photons in this energy range is today more like old-fashioned naked-eye observations than to modern device-assisted with an optical tool, because of the impossibility of concentrating such high-energy photons. Indeed, the lack of optical components working in this energy range results in the impossibility of focusing, which in turn means poor signal-to-noise ratio recorded by the detectors.

One nonfocusing method that has already been proposed for X-ray detection consists in the usage of geometrical optics, such as collimators or coded masks \cite{UbertiniIntegral}. However, since the total interaction cross-section for $\gamma$-rays attains its minimum within 100-1000 keV, the efficiency of geometrical optics decreases while at the same time the background noise increases with respect to the signal, because of the growing importance of shield leakage and/or n$\beta$ activation. Another nonfocusing solution consists of quantum optics based on Compton effect and tracking detectors \cite{PeterAstro}.

Focusing methods have greater potential because they can concentrate the signal from a large collector onto a small detector and beat the instrumental background that may hamper the observation. Bragg diffraction can be used to concentrate with high efficiency the signal. As focusing optics, multilayers have been proven to be capable of focusing of up to 80 keV photons with high-efficiency \cite{MadsenSPIE}. More recently it was demonstrated that multilayer reflective optics could operate efficiently and according to classical-wave physics up to photon energies of at least 384 keV \cite{multilayerprl}. However, these new reflective optics work at very low grazing incidence angles, below $0.1^°$, thus featuring a very low acceptance area for the incident photons, and beyond these energy limits their efficiency critically deteriorates.

\begin{figure}
\begin{center}
  \includegraphics[width=0.85\textwidth]{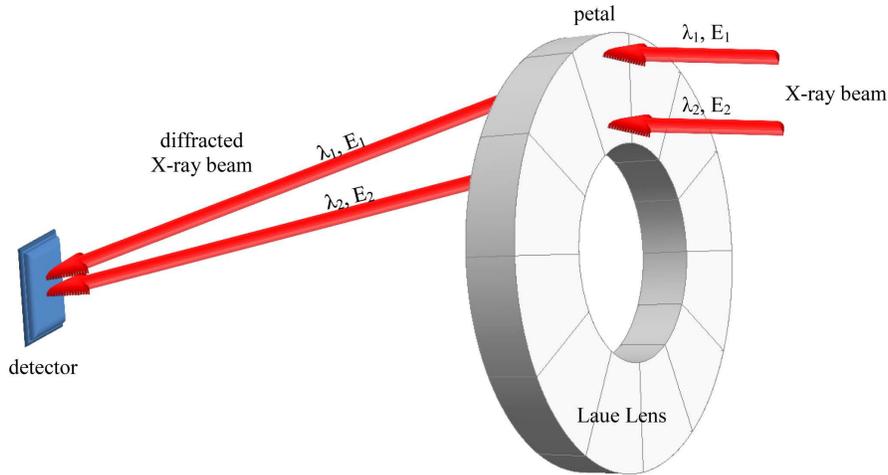}
  \caption{Schematic representation of a Laue lens. The red arrows represent an X-ray beam that undergoes diffraction toward a detector placed on the focal point of the lens.}\label{fig:lens}
  \end{center}
\end{figure}

If X-ray diffraction occurs traversing the crystal (the Laue scheme), the problem of focusing hard X-rays can be approached via a Laue lens \cite{Lund}. A Laue lens is conceived as an ensemble of many crystals arranged in such a way that as much radiation as possible is diffracted onto the lens focus over a selected energy band (see Fig. \ref{fig:lens}). A Laue lens can be employed for astrophysical purposes for high-sensitivity observations of many cosmic phenomena. Indeed, X- and $\gamma$-ray emissions take place in several places in the Universe, from near solar flares, to compact binary systems, pulsars and supernova remnants within our Galaxy, and finally to extremely distant objects such as active galactic nuclei and $\gamma$-ray bursts (GRBs) at redshifts z $>$ 8.

A Laue lens could also be used for high-quality diagnosis in nuclear medicine. For instance, it would improve $\gamma$-ray detection in single-photon emission computed tomography (SPECT) and in positron emission tomography (PET) by providing better scan resolution. This, in turn, would lead to a lower radioactive dose being imparted to the patient, since there would be no need for tomography scanning \cite{Roa.medica}.

A Laue lens can also be used for concentrate hard X-rays for therapy purposes. Indeed, radiation therapy uses high-energy radiation to kill cancer cells by damaging their DNA. Since the radiation therapy can damage normal cells as well as cancer cells, the treatment must be carefully planned to minimize side effects. With a Laue lens it would be possible to minimize the collateral effects of the radiation therapy and to improve its functionality.

Diffracting crystals are the optical elements of a Laue lens. To date, several materials have been proposed for diffracting X- and $\gamma$-rays. The first proposals date back more than 30 years ago \cite{Smitherlens}. Several steps forward have been accomplished, both in the growth of high-quality crystals and in their processing.

In order to build a high-efficiency Laue lens, the choice and the arrangement of the crystals are key points. Since the implementation of a Laue lens may be very expensive, an appropriate simulation of its features before its realization is fundamental. However, the optimized arrangement of the crystals on a Laue lens is not approachable analytically. Indeed, a genetic algorithm could be the best method to resolve the issue. A genetic algorithm is a search heuristic that mimics the process of natural selection. With a genetic algorithm, a population of candidate solutions to an optimization problem is evolved toward better solutions. In particular, a Laue lens can be optimized through a pseudo-evolutionistic process able to favor the best crystals and thus bring the system towards a configuration with the desired features.

In this paper the \emph{LaueGen} genetic algorithm is proposed. \emph{LaueGen} is capable of selecting and arranging the crystal tiles in a Laue lens in such a way to maximize its integrated reflectivity together with smoothing the energy dependence of the collected photons. Smoothing the spectral response is important to simplify the deconvolution of the collected signal, while the high-reflectivity of the optical elements is fundamental for increasing the signal-to-noise ratio of the lens.

In next section the description and the theoretical reflectivity of the crystals proposed so far for a Laue lens are reported. Then, the code is described, and finally some examples are given.

\section{Crystals for high-efficiency diffraction}\label{sec:cry}

To realize a Laue lens, crystals that can diffract the radiation with high efficiency and with a controlled passband response are needed. For this purpose, the scientific community has proposed several kinds of crystals, namely mosaic crystals \cite{zach} and crystals with curved diffracting planes (CDP) \cite{authier.2001fe}. More recently, a particular optical element based on CDPs has been proposed, namely the quasi-mosaic (QM) crystal, which is a crystal with two different curvatures in two lyings of perpendicular planes \cite{Sumbaev.quasimosaicity}. The features of such crystals are quite different, which will be described in the next subsections.

\subsection{Mosaic crystals}

A mosaic crystal can be described using Darwin's model as an assembly of tiny identical small perfect crystals, called crystallites, each slightly misaligned with respect to the others according to an angular distribution, usually taken as Gaussian, spread about a nominal direction \cite{zach}. The presence of misaligned crystallites provides an enlarged bandpass for the diffracted photons. For a mosaic crystal, reflectivity is given by \cite{Barriere:he5432}
\begin{equation}
\label{eq:refmosaic}
    \eta_M = \frac{1}{2}(1-e^{-2W(\Delta\theta)QT_0})e^{\frac{-\mu T_0}{cos \theta _B}}
\end{equation}
where $T_0$ is the crystal thickness traversed by radiation, $\Delta \theta$ the difference between the angle of incidence and the Bragg angle $\theta_B$, $\mu$ the linear absorption coefficient within the crystal, and $W(\Delta\theta)$ the distribution function of crystallite orientations. In turn, $W(\Delta\theta)$ is defined as
\begin{equation}
W(\Delta \theta)= 2\bigg(\frac{ln2}{\pi}\bigg)^\frac{1}{2}\frac{1}{\Omega_M}e^{-ln2(\frac{\Delta\theta}{\Omega_M/2})^2}
\end{equation}
where $\Omega_M$ is called mosaicity, or mosaic spread, and represents the full width at half maximum (FWHM) of the angular distribution of the crystallites. Finally, by considering the kinematical theory approximation \cite{malgrange.2002zm}, $Q$ is given by
\begin{equation}
   Q=\frac{\pi^2d_{hkl}}{\Lambda^2_0cos \theta _B}
\end{equation}
where $d_{hkl}$ is the d-spacing of the diffracting planes, and $\Lambda_0$ the extinction length as defined in \cite{authier.2001fe} for the Laue symmetric case.

\subsection{CDP crystals}

Perfect crystals with curved diffracting planes (CDP crystals) represent an alternative to mosaic crystals. The curvature of the diffracting planes results in a very well controlled of the energy bandpass, because it is proportional to the angular distribution of the diffracting planes. For CDP crystals, the reflectivity is given by \cite{malgrange.2002zm}
\begin{equation}
\label{eq:refc}
    \eta_{C} = (1-e^{\frac{-\pi^2T_0d_{hkl}}{\Omega_C\Lambda^2_0}})e^{\frac{-\mu T_0}{cos \theta _B}}
\end{equation}
where, in this case, $\Omega_C$ represents the bending angle of the curved diffracting planes. The angular distribution of the diffracting planes $W_C(\Delta\theta)$ turns out to be a uniform distribution with the width being the angular spread $\Omega_C$. It is
\begin{eqnarray}
{W_C(\Delta\theta)=UnitBox(\Omega_{C})=}\nonumber\\
{\left\{ \begin{array}{ll}
 1 &\mbox{ for } -\frac{\Omega_C}{2} < \theta-\theta_B < \frac{\Omega_C}{2}\\
 0 &\mbox{ otherwise}
       \end{array} \right.}
\end{eqnarray}

\subsection{QM crystals}

A QM crystal features two curvatures of two different lying of crystallographic planes. Indeed, as a crystal is bent by external forces, under very specific orientations another curvature can be generated within the crystal, namely the QM curvature. Due to the external curvature, this kind of crystal permits the focalization of the radiation on the scattering plane \cite{QM}. Since QM crystals belong to the CDP class, their reflectivity is given by Eq.\ref{eq:refc}. However, the angular distribution of the diffracting planes is the convolution between the external and the QM curvature, namely

\begin{eqnarray}
{W_{QM}(\Delta\theta)=\frac{UnitBox(\Omega_{C})}{\Omega_{C}}\ast UnitBox(\Omega_{QM})=}\nonumber\\
{\left\{ \begin{array}{ll}
 \frac{\Omega_{QM}}{\Omega_C} &\mbox{ for }
 -\frac{\Omega_C-\Omega_{QM}}{2} < \theta-\theta_B < \frac{\Omega_C-\Omega_{QM}}{2}\\
 \frac{\Omega_C+\Omega_{QM}+2\theta}{2\Omega_C} &\mbox{ for }
 -\frac{\Omega_C+\Omega_{QM}}{2} < \theta-\theta_B < -\frac{\Omega_C-\Omega_{QM}}{2}\\
 \frac{\Omega_C+\Omega_{QM}-2\theta}{2\Omega_C} &\mbox{ for }
 \frac{\Omega_C-\Omega_{QM}}{2} < \theta-\theta_B < \frac{\Omega_C+\Omega_{QM}}{2}\\
 0 &\mbox{ otherwise}
       \end{array} \right.}
\end{eqnarray}

Some examples of QM crystals employed for high-focusing diffraction are reported in \cite{CamaQM,LaueQM}.

\section{The \emph{LaueGen algorithm}}\label{sec:LaueGen}

The particular characteristics of a Laue lens may vary substantially according to the application which the lens is needed for. In order to be general and highly versatile, the program allows the user to set several initial parameters. These parameters are
\begin{itemize}
  \item the focal length $f$;
  \item the energy band;
  \item the size of the crystals;
  \item the minimum and the maximum radius of the Laue lens $R_{min}$ and $R_{max}$;
  \item the material of the crystals (including the kind, e.g., mosaic, CDP or QM);
  \item the lying of diffracting planes.
\end{itemize}

The focal length of a Laue lens for astrophysical purposes can change depending on the structure in which the lens has to be mounted. For example, in the case of a balloon, the focal length must not exceed a few meters \cite{ballmoos.2005ek}. If the lens is launched through a satellite, the focal length may be up to 20 m \cite{SPIE2012Laue,SPIE2013Dante}. Finally, in case of two satellites flying in formation, the focal length may be up to 100 m \cite{MAX,dual}. On the contrary, the focal length of a Laue lens designed for nuclear medicine applications is limited to 10-50 cm \cite{smither.2005la}.

The diameter of a Laue lens depends on the energy band that it has to cover and on the crystals chosen as optical elements. Indeed, the distance $R$ from the axis of the lens at which a crystal diffracts the radiation onto the detector is proportional to the Bragg angle, i.e., $R$ depends on the crystallographic planes used for diffraction. By using the small-angle approximation, it is
\begin{equation}
    R=f \tan(2\theta_B) \propto \sqrt{h^2+k^2+l^2}
\end{equation}
where $(h,k,l)$ are the Miller indices of the planes used for diffraction. The diffraction efficiency for planes with high Miller indices is lower than for the planes with small indices. However, their external position in the Laue lens guarantees a large geometric area, resulting in a large effective area. The effective area of a Laue lens at a certain energy is defined as the geometric area of the lens, as seen by the X-ray beam, times the diffraction reflectivity of the crystals at the energy of interest. Such parameter is important to quantify the number of events that an ideal detector located on the focus of the Laue lens would count under exposure to a given photon flux.

The selection of the energy range of the Laue lens depends on the purpose of the lens itself. Accordingly, the user has to select the crystals that the code has to take into account. Then, during the run, \emph{LaueGen} calculates which material has to be used and how to arrange the different crystals. The convergence speed of the code depends on the degrees of freedom with which the code is initialized.

Once the parameters have been set, the code begins running. Firstly, the number of crystals for each ring is calculated as the integer part of
\begin{equation}\label{eq:N}
\frac{\pi}{ArcTan(\frac{L_{tan}+B}{2R})}
\end{equation}
where $B$ is the minimum distance between two neighboring samples, needed to prevent that they touch each others. $L_{tan}$ is the side length of the samples along the direction perpendicular to the ring radius, i.e. the tangential direction.

The angular spread of the diffracting planes $\Omega$ is calculated to be proportional to the radial size of the tiles $L_{rad}$, in order to prevent the formation of voids between two neighboring rings in the effective area
\begin{equation}\label{eq:spread}
\Omega=\frac{L_{rad}+B}{2f}
\end{equation}
The thickness $T_0$ of each crystals is optimized as a function of the energy and the angular spread of the diffracting planes, following the procedure reported in \cite{Barriere:he5432}.

Then, for each kind of crystal, the effective area is computed, taking into account every ring that constitute the lens. The effective area $A_{Eff}$ for a single crystal is
\begin{equation}
A_{Eff}(E)_{(tile)} = \eta(E) A_{tile}
\end{equation}
where $\eta(E)$ is the reflectivity of the crystals as a function of the energy of the impinging beam and $A_{tile}$ is the geometric area of each tile. The effective area of the whole lens is the sum of the effective areas of the tiles composing the lens itself. The effective area is calculated in the code as a discrete function of the energy, with the step $\Delta E$ being selected as a tradeoff between accuracy and computational time.

As an estimator of the smoothness of the effective area as a function of the photon energy, the quantity $S_{Eff}$ is defined in the following way. Firstly, the moving average $ma_{Eff}$ of $A_{Eff}$ is calculated for $n$ sequential steps of energy, by starting from the minimum.
\begin{equation}
ma_{Eff}(k)=\frac{\sum_{j=0}^n A_{Eff}(E_{min}+(j+k)\Delta E)}{n}
\end{equation}
$k$ represents the position of the window of the moving average. At the first step, $k=0$. Then, the distance between the values of $A_{Eff}$ to their average value $ma_{Eff}(k)$ is calculated and the largest one is being selected.
\begin{eqnarray}
S(k)=max\biggl(|ma_{Eff}(k)-A_{Eff}(E_{min}+k\Delta E)|,\nonumber \\ |ma_{Eff}(k)-A_{Eff}(E_{min}+(k+1)\Delta E)|,...,\nonumber \\ |ma_{Eff}(k)-A_{Eff}(E_{min}+(k+n)\Delta E)|\biggl)
\end{eqnarray}
This procedure has to be repeated by shifting the window under analysis along the energy axis, up to the maximum energy, i.e., by starting from $k=0$ to $k=E_{max}-E_{min}-n\Delta E$. Then, the quantity $S_{Eff}$ is the sum of all the collected contributions of maximum distances
\begin{equation}
S_{Eff}=\sum_{k=0}^{E_{max}-E_{min}-n\Delta E} S(k)
\end{equation}

At this stage, the Laue lens can be initialized, filling the rings with the selected crystals for the simulation. The filling can be performed either randomly or with an \emph{a priori} initial guess for the tile arrangement.

To simultaneously maximize and smoothen the effective area of a simulated Laue lens, the code operate through the following genetic algorithm. It may be considered as an evolution of the code described in \cite{Pisaphd}. A tile of the lens is chosen casually and randomly transformed into a tile of different species by changing the material and/or the crystallographic orientation. Then, the new configuration is evaluated in terms of the integrated reflectivity of the Laue lens, i.e., the integral of its effective area, and the smoothness of the effective area itself as a function of energy. A control parameter that can evaluate the effective area of the lens and its dependence on energy can be written as a function of $A_{Eff}$ and $S_{Eff}$ in this way
\begin{equation} \label{eq:genetic}
k[i]=w_1\frac{\int_{E_{min}}^{E_{max}} A_{Eff}(E)[i+1]\, dE}{\int_{E_{min}}^{E_{max}} A_{Eff}(E)[i]\, dE}-w_2\frac{S_{Eff}[i+1]}{S_{Eff}[i]}
\end{equation}
where the index [\emph{i}] represents the configuration before the crystal transformation, while the index [\emph{i+1}] represents the configuration after the crystal transformation. $w_1$ and $w_2$ are weights to be assigned to the two terms in Eq.\ref{eq:genetic}. By increasing $w_1$ it is possible to privilege a high value of the total effective area, while an increase of $w_2$ favors a better smoothness. The sign minus before the second term indicates that the second quantity has to be minimized, while the first has to be maximized.

To speed the calculation, the integral $\int_{E_{min}}^{E_{max}} A_{Eff}[i]\, dE$ can be calculated only one time at the beginning of the genetic algorithm. The term $[i+1]$ is
\begin{eqnarray}
\int_{E_{min}}^{E_{max}} A_{Eff}[i+1]\, dE= \nonumber\\ \int_{E_{min}}^{E_{max}} A_{Eff}[i]\, dE-A_{Eff}(E)_{tile(i)}+A_{Eff}(E)_{tile(i+1)}
\end{eqnarray}
where the subscript $tile(i)$ signifies the contribution to the effective area of the removed crystal, while $tile(i+1)$ the contribute of the added crystal.

If $k[i+1]>k[i]$, the crystal exchange is held else it is rejected. The program runs until the system reaches the thermalization, i.e., once the maximum of $k[i]$ has been attained.

\section{Example}\label{sec:ex}

An example of the output of the code is reported. The program was initialized to consider the 500-850 keV energy range. This energy range was chosen because it represents a window of great interest for an astrophysical mission. Indeed, the study of the origin of the positrons annihilating in the Galactic center could be visible through the e$^+$ / e$^-$ annihilation line at 511 keV. A study of the distribution of this emission line would thus bring new clues concerning the still elusive sources of antimatter. Another significant observation is the 847 keV line, produced in the decay of $^{56}$Co nuclei in Supernovae Ia, which is a $\gamma$-ray line of highest astrophysical relevance.

As optical elements, germanium, silicon, and copper were chosen, because they are the most experimented crystals in sight of building a Laue lens for hard X-ray diffraction. According with the literature, Ge and Si were chosen to be CDP crystals, Cu was chosen as a mosaic crystal. (111) and (220) crystallographic planes were chosen because they highlight the most intense efficiencies for diffraction. A focal length $f$ equal to 20 m was selected because it fits the case of an astrophysical mission based on a satellite.

The effective area was calculated as a discrete function of the energy, the step being 1 keV. This was a good compromise to obtain a precise calculation of the effective area together with saving computational time. All the initial parameters are showed in Tab.\ref{tab:parametersLauelGen}

\begin{table}
\begin{center}
\caption{Initial parameters of the simulated Laue lens}\label{tab:parametersLauelGen}
\begin{tabular}{rc}
\hline
focal length (m) & 20\\
inner ring radius (m) & 0.085\\
outer ring radius (m) & 0.25\\
CDP crystals & Ge and Si\\
mosaic crystals & Cu \\
crystallographic planes & (111), (220)\\
crystals size (mm$^2$) &10$\times$10\\
min samples distance $B$ (mm) & 2\\
energy range (keV) & 500-850\\
energy step discretization $\Delta E$ (keV) & 1\\
window of the moving average (n) & 100\\
$w_1$ - $w_2$ &1 - 100\\
\hline
\end{tabular}
\end{center}
\end{table}

At this stage, the code calculated the number of crystals through Eq.\ref{eq:N} and the angular spread of the samples through Eq.\ref{eq:spread}. This latter turned out to be 56.72 arcsec.

\begin{figure}
\begin{center}
  \includegraphics[width=0.75\textwidth]{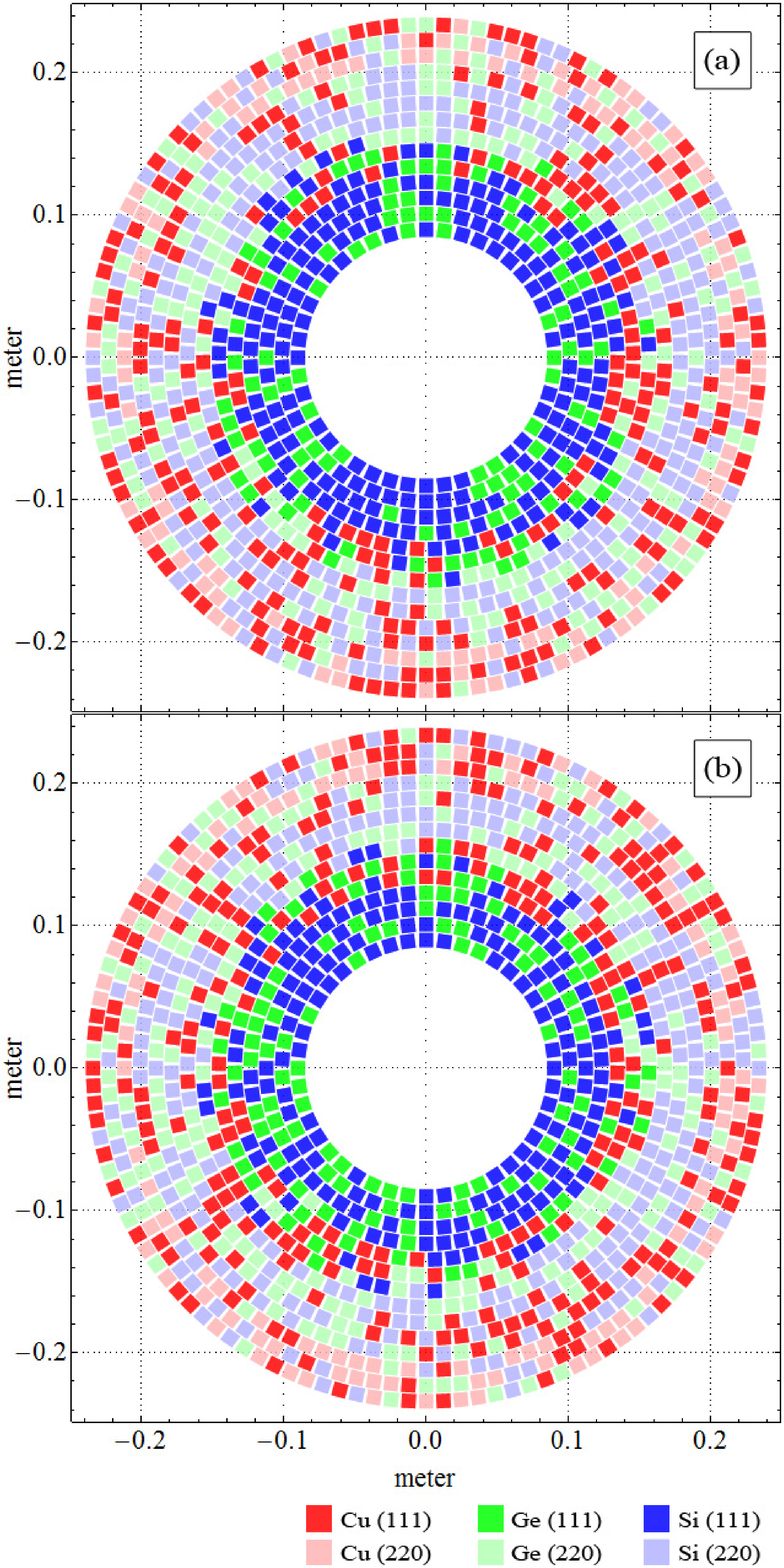}
  \caption{Schematic representation of the simulated Laue lens. a) Code initialized with only Ge (111) crystals. b) Code initialized with random crystals disposition.}\label{fig:lensLaueGen}
  \end{center}
\end{figure}

\begin{figure}
\begin{center}
  \includegraphics[width=0.75\textwidth]{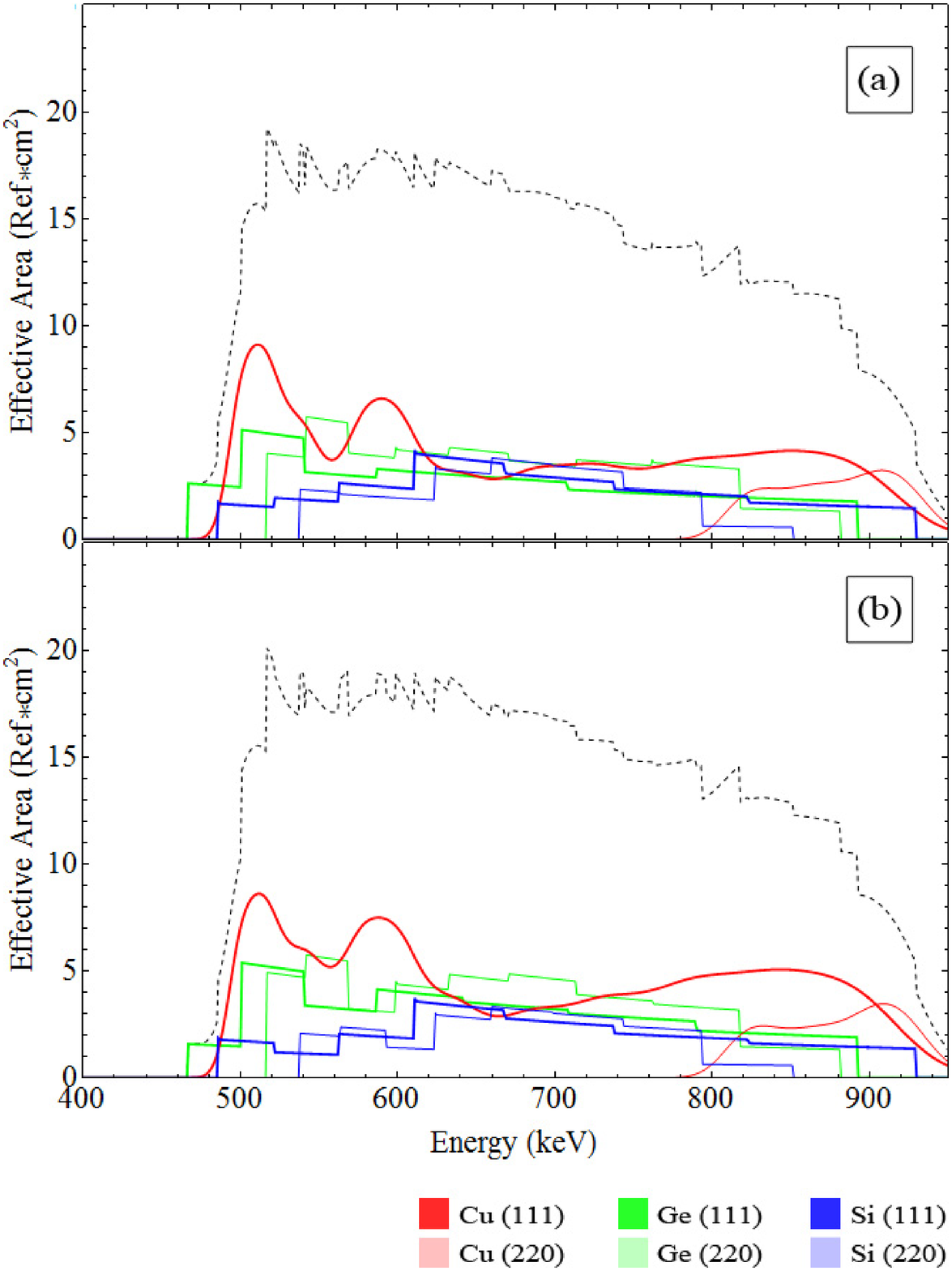}
  \caption{Effective area per unit energy of the lens. The contributions of the tiles of different species are visible. Tiles were positioned in the lens to maximize effective area in the energy range 500-850 keV and made the profile as smooth as possible with respect to energy variation. The dashed line is the total effective area. a) Code initialized with only Ge (111) crystals. b) Code initialized with a random crystal disposition.}\label{fig:AE}
  \end{center}
\end{figure}

The code was cycled until the system attained the thermalization. In order to verify that the final arrangement was not affected by the initial guess and to control possible interferences of local maxima in the quantifier of Eq.\ref{eq:genetic}, the algorithm was initialized with two different dispositions of the tiles. Firstly, it was initialized with only Ge (111) CDP crystals, then with random crystals, chosen among the crystals selected at the beginning. In Fig.\ref{fig:lensLaueGen} are shown the final arrangements of the crystals for the two cases, while Fig.\ref{fig:AE} shows the effective area of the simulated Laue lens for the two cases. Finally, in Tab.\ref{tab:AE} are listed the crystals disposition obtained.

\begin{table}
\caption{Crystals composing the simulated Laue lens. The first numbers in the columns represent the amount of crystals for the Laue lens initialized with only Ge crystals, while the second numbers concern the random initialization case.}\label{tab:AE}
\begin{center}
\begin{tabular}{lcccccc}
\hline
ring &Ge(111)&Ge(220)&Cu(111)&Cu(220)&Si(111)&Si(220)\\
\hline
1 & 14 - 15 & 0 - 0 & 0 - 0 & 0 - 0 & 30 - 29 & 0 - 0 \\
2 & 15 - 18 & 0 - 0 & 0 - 0 & 0 - 0 & 35 - 32 & 0 - 0 \\
3 & 16 - 20 & 0 - 0 & 0 - 0 & 0 - 0 & 40 - 36 & 0 - 0 \\
4 & 16 - 20 & 0 - 1 & 0 - 0 & 0 - 0 & 46 - 41 & 0 - 0 \\
5 & 14 - 15 & 0 - 0 & 26 - 31 & 0 - 0 & 27 - 21 & 0 - 0 \\
6 & 21 - 22 & 11 - 11 & 23 - 29 & 0 - 0 & 18 - 11 & 0 - 0 \\
7 & 10 - 6 & 25 - 24 & 18 - 22 & 0 - 0 & 14 - 15 & 12 - 12 \\
8 & 0 - 0 & 24 - 25 & 19 - 18 & 0 - 0 & 0 - 0 & 42 - 42 \\
9 & 0 - 0 & 22 - 29 & 14 - 14 & 0 - 0 & 0 - 0 & 55 - 48 \\
10 & 0 - 0 & 24 - 27 & 16 - 20 & 0 - 0 & 0 - 0 & 56 - 49 \\
11 & 0 - 0 & 22 - 23 & 35 - 39 & 0 - 0 & 0 - 0 & 45 - 40 \\
12 & 0 - 0 & 20 - 16 & 16 - 23 & 47 - 51 & 0 - 0 & 25 - 18 \\
13 & 0 - 0 & 27 - 27 & 29 - 29 & 33 - 30 & 0 - 0 & 25 - 28 \\
14 & 0 - 0 & 18 - 22 & 46 - 43 & 30 - 31 & 0 - 0 & 26 - 24 \\
tot & 106 - 116 & 193 - 205 & 242 - 268 & 110 - 112 & 210 - 185 & 286 - 261 \\
\hline
\end{tabular}
\end{center}
\end{table}

\begin{figure}
\begin{center}
  \includegraphics[width=0.6\textwidth]{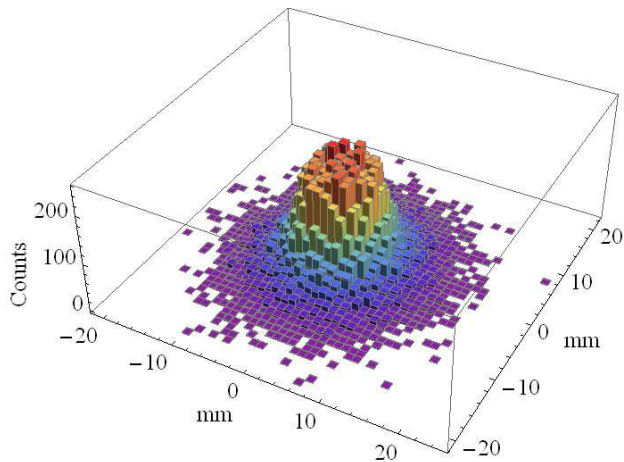}
  \caption{Photon distribution onto the focal plane of the lens in arbitrary units. Concentration of photons is high in the focal point and rapidly decreases far from the center. The distribution is obtained with a Monte Carlo simulation.}\label{fig:fotoni}
  \end{center}
\end{figure}

After the two tests, Figs. \ref{fig:AE} and \ref{fig:lensLaueGen} highlight an equivalent arrangement of the crystal tiles. Indeed, although the number of crystals per type and per ring are not exactly the same, very similar effective areas have been obtained, which resulted smooth and high for both cases. In Fig. \ref{fig:fotoni} the photon distribution on the focal plane is shown. It can be noticed that the concentration of photons is high at the center of the spot and rapidly decreases farther from the center.

Other examples of the use of \emph{LaueGen} can be found in \cite{AA,311}, where the code has been used to design very large Laue lens, entirely based on quasi-mosaic crystals.

\section{Conclusions}

The \emph{LaueGen} code has been implemented for the design of optimized Laue lenses. The program is based on a genetic algorithm. It can take into account any kind of diffracting crystals, and combining them in order to obtain the best results in terms of integrated reflectivity. In particular, the user can decide whether to prefer a lens with an high effective area or a smooth spectral response of the diffracted photons. Moreover, it has been shown that the final configuration of the crystals in the lens does not depend on the initial guess of initialization, i.e. the global maximum can be attained.

\emph{LaueGen} can generate lens configurations for very different applications, spanning from astrophysical to medical purposes. The code has been proven to work with an example shown in this paper and with other two examples shown in the literature \cite{AA,311}. Finally, the code could be a valuable tool for the design of future experiments based on Laue lens.

\section{acknowledgements}
The authors are thankful to INFN for financial support through the LOGOS project.

\bibliographystyle{unsrt}
\bibliography{biblio}

\end{document}